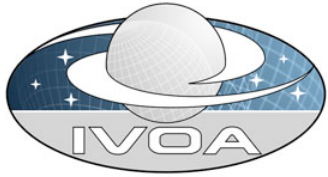

# Vocabularies in the Virtual Observatory
# Version 1.19

## IVOA Recommendation, 2009 October 7

**Editors**
    Alasdair J G Gray, University of Manchester, UK
    Norman Gray, University of Leicester / University of Glasgow, UK
    Frederic V Hessman, University of Göttingen, Germany
    Andrea Preite Martinez, INAF, Italy
**Authors**
    Sébastien Derriere, Alasdair J G Gray, Norman Gray, Frederic V Hessman, Tony Linde, Andrea Preite Martinez, Rob Seaman and Brian
    Thomas

---

## Abstract

This document specifies a standard format for vocabularies based on the W3C's *Resource Description Framework* (RDF) and *Simple Knowledge Organization System* (SKOS). By adopting a standard and simple format, the IVOA will permit different groups to create and maintain their own specialised vocabularies while letting the rest of the astronomical community access, use, and combine them. The use of current, open standards ensures that VO applications will be able to tap into resources of the growing semantic web. The document provides several examples of useful astronomical vocabularies.

## Status of this document

This is an IVOA Recommendation.

The first release of this document was 2008 March 20. It has been reviewed by IVOA Members and other interested parties, and has been endorsed by the IVOA Executive Committee as an IVOA Recommendation. It is a stable document and may be used as reference material or cited as a normative reference from another document. IVOA's role in making the Recommendation is to draw attention to the specification and to promote its widespread deployment. This enhances the functionality and interoperability inside the Astronomical Community.

A list of current IVOA Recommendations and other technical documents can be found at http://www.ivoa.net/Documents/.

## Note on conformance

Text within the following document is classified as either "normative" or "informative".

**Normative** text means information that is required to implement the Recommendation; an implementation of this Recommendation is conformant if it abides by all the prescriptions contained in normative text. **Informative** text is information provided to clarify or illustrate a requirement but which is not required for conformance.

The sections and subsections of this Recommendation are labeled, after the section heading, to specify whether they are normative or informative. If a subsection is not labeled, it has the same normativity as its parent section. References are normative if they are referred to within normative text.

Within normative sections, the key words MUST, MUST NOT, REQUIRED, SHALL, SHALL NOT, SHOULD, SHOULD NOT, RECOMMENDED, MAY, and OPTIONAL in this document are to be interpreted as described in *[std:rfc2119]*.

## Acknowledgments

We would like to thank the members of the IVOA semantic working group for many interesting ideas and fruitful discussions, and Antoine Isaac of the W3C SKOS WG for notes on SKOS conformance.

## Table of Contents

---

# 1. Introduction (informative)

## 1.1. Vocabularies in astronomy

Astronomical information of relevance to the Virtual Observatory (VO) is not confined to quantities easily expressed in a catalogue or a table. Fairly simple things such as position on the sky, brightness in some units, times measured in some frame, redshifts, classifications or other similar quantities are easily manipulated and stored in VOTables and can currently be identified using IVOA Unified Content Descriptors (UCDs) [std:ucd]. However, astrophysical concepts and quantities use a wide variety of names, identifications, classifications and associations, most of which cannot be described or labelled via UCDs.

There are several basic forms of organised semantic knowledge of potential use to the VO. Informal "folksonomies" are at one extreme, and are a very lightly coordinated collection of labels chosen by users. A slightly more formal structure is a "vocabulary", where the label is drawn from a predefined set of definitions which can include relationships to other labels; vocabularies are primarily associated with searching and browsing tasks. At the other extreme are "ontologies", where the domain is formally captured in a set of logical classes, typically related in a subclass hierarchy. More formal definitions are presented later in this document.

An astronomical ontology is necessary if we are to have a computer (appear to) "understand" something of the domain. There has been some progress towards creating an ontology of astronomical object types [std:ivoa-astro-onto] to meet this need. However there are distinct use cases for letting human users find resources of interest through search and navigation of the information space. The most appropriate technology to meet these use cases derives from the Information Science community, namely that of *controlled vocabularies, taxonomies and thesauri*. In the present document, we do not distinguish between controlled vocabularies, taxonomies and thesauri, and use the term *vocabulary* to represent all three.

One of the best examples of the need for a simple vocabulary within the VO is VOEvent [std:voevent], the VO standard for supporting rapid notification of astronomical events. This standard requires some formalised indication of what a published event is "about", in a formalism which can be used straightforwardly by the developer of relevant services. See 1.2. Use-cases, and the motivation for formalised vocabularies for further discussion.

A number of astronomical vocabularies have been created, with a variety of goals and intended uses. Some examples are detailed below.

- The *Second Reference Dictionary of the Nomenclature of Celestial Objects* [lortet94], [lortet94a] contains 500 paper pages of astronomical nomenclature

- For decades professional journals have used a set of reasonably compatible keywords to help classify the content of whole articles. These keywords have been analysed by Preite Martinez & Lesteven [preitemartinez07], who derived a set of common keywords constituting one of the potential bases for a fuller VO vocabulary. The same authors also attempted to derive a set of common concepts by analysing the contents of abstracts in journal articles, which should comprise a list of tokens/concepts more up-to-date than the old list of journal keywords. A similar but less formal attempt was made by Hessman [hessman05] for the VOEvent working group, resulting in a similar list.

- Astronomical databases generally use simple sets of keywords – sometimes hierarchically organised – to help users make queries. Two examples from very different contexts are the list of object types used in the Simbad database and the search keywords used in the educational Hands-On Universe image database portal.

- The Astronomical Outreach Imagery (AOI) working group has created a simple taxonomy for helping to classify images used for educational or public relations [std:avm]. See 4.2. The AVM Taxonomy.

- In 1993, Shobbrook and Shobbrook published an Astronomy Thesaurus endorsed by the IAU [shobbrook93]. This collection of nearly 3000 terms, in five languages, is a valuable resource, but has seen little use in recent years. Its very size, which gives it expressive power, is a disadvantage to the extent that it is consequently hard to use. See 4.4. The 1993 IAU Thesaurus.

- The VO's Unified Content Descriptors [std:ucd] (UCD) constitute the main controlled vocabulary of the IVOA and contain some taxonomic information. However, UCD has some features which supports its goals, but which make it difficult to use beyond the present applications of labelling VOTables: firstly, there is no standard means of identifying and processing the content of the text-based reference document; secondly, the content cannot be openly extended beyond that set by a formal IVOA committee without going through a laborious and time-consuming negotiation process of extending the primary vocabulary itself; and thirdly, the UCD vocabulary is primarily concerned with data types and their processing, and only peripherally with astronomical objects (for example, it defines formal labels for RA, flux, and bandpass, but does not mention the Sun). See 4.3. The UCD1+ Vocabulary.

## 1.2. Use-cases, and the motivation for formalised vocabularies

The most immediate high-level motivation for this work is the requirement of the VOEvent standard [*std:voevent*] for a controlled vocabulary usable in the VOEvent's `<Why/>` and `<What/>` elements, which describe what sort of object the VOEvent packet is describing, in some broadly intelligible way. For example a "burst" might be a gamma-ray burst due to the collapse of a star in a distant galaxy, a solar flare, or the brightening of a stellar or AGN accretion disk, and having an explicit list of vocabulary terms can help guide the event publisher into using a term which will be usefully precise for the event's consumers. A free-text label can help here (which brings us into the domain sometimes referred to as folksonomies), but the astronomical community, with a culture sympathetic to international agreement, can do better.

This standard establishes a set of conventions for the creation, publication, use, and manipulation of astronomical vocabularies within the Virtual Observatory, based upon the W3C's SKOS standard. We include as appendices to this standard formalised versions of a number of existing vocabularies, encoded as SKOS vocabularies [*std:skosref*].

Specific use-cases include the following.

- A user wishes to process all events concerning supernovae, which means that an event concerning a type 1a supernova must be understood to be relevant. [This supports a system working autonomously, filtering incoming information.]

- A user is searching an archive of VOEvents for microlensing events, and retrieves a large number of them; the search interface may then prompt them to narrow their search using one of a set of terms including, say, binary lens events. [This supports so-called "semantic search", providing semantic support to an interface which is in turn supporting a user.]

- A user wishes to search for resources based on the journal-supplied keywords in a paper; they might either initiate this by hand, or have this done on their behalf by a tool which can extract the keywords from a PDF. The keywords are in the A&A vocabulary, and mappings have been defined between this vocabulary and others, which means that the query keywords are translated automatically into those appropriate for a search of an outreach image database (everyone likes pretty pictures), the VO Registry, a set of Simbad object types, and one or more concepts in more formal ontologies. The search interface is then able to support the user browsing up and down the AVM vocabulary, and a specialised Simbad tool is able to take over the search, now it has an appropriate starting place. [This supports interoperability, building on the investments which institutions and users have made in existing vocabularies.]

- A user receives a VOTable of results from a VO application – for example a catalogue of objects or observations – and wants to search a database of old FITS files for potential matches. Because the UCDs labeling the columns of the tables are expressed in well-documented SKOS, both the official descriptions of the UCDs and their semantic matches to a variety of other plain-text vocabularies (such as the IAU or AVM thesauri) are available to the VO application, providing a basis for massive searches for all kinds of FITS keyword values.

The goal of this standard is to show how vocabularies can be easily expressed in an interoperable and computer-manipulable format, and the first normative section of this Recommendation (namely section 3. Publishing vocabularies (normative)) contains requirements and suggestions intended to promote this. In the other normative section (4. SKOS versions of existing vocabularies (normative)), we include four example vocabularies that have previously been expressed using non-standardized formats – namely the A&A keyword list, the IAU thesaurus and AVM taxonomy, and UCD1. These are included as illustrations of how simple it is to publish them in SKOS, without losing any of the information of the original source vocabularies.

It is not a goal of this standard, as it is not a goal of SKOS, to produce knowledge-engineering artefacts which can support elaborate machine reasoning – the use-cases above are broadly concerned with searching for data, rather than representing the data itself, and for this the looser semantics of a thesaurus are more appropriate than the formal ("is-a") semantics of an ontology. More elaborate artefacts would be very valuable, but require much more expensive work on ontologies. As the supernova use-case above illustrates, even simple vocabularies can support useful machine reasoning.

It is also not a goal of this standard to produce new vocabularies, or substantially alter existing ones; instead, the vocabularies included below in section 4. SKOS versions of existing vocabularies (normative) are directly and mechanically derived from existing vocabularies (the exceptions are the IVOAT vocabulary, which is ultimately intended to be a significant update to the IAU-93 original, and the constellations vocabulary, which is intended to be purely didactic). It therefore follows that the ambiguities, redundancies and incompleteness of the source vocabularies are faithfully represented in the distributed SKOS vocabularies. We hope that this formalisation process will create greater visibility and broader use for the various vocabularies, and that this will guide the maintenance efforts of the curating groups.

The reason for both of these limitations is that vocabularies are extremely expensive to produce, maintain and deploy, and we must therefore rely on such vocabularies as have been developed, and attached as metadata to resources, by others. Such vocabularies are less rich or less coherent than we might prefer, but they are widely enough deployed to be useful. We hope that the set of example vocabularies we have provided will build on this deployment, by providing material which is useful out of the box.

## 1.3. Formalising and managing multiple vocabularies

We find ourselves in the situation where there are multiple vocabularies in use, describing a broad range of resources of interest to professional and amateur astronomers, and members of the public. These different vocabularies use different terms and different relationships to support the different constituencies they cater for. For example, "delta Sct" and "RR Lyr" are terms one would find in a vocabulary aimed at professional astronomers, associated with the notion of "variable star"; however one would *not* find such technical terms in a vocabulary intended to support outreach activities.

One approach to this problem is to create a single consensus vocabulary, which draws terms from the various existing vocabularies to create a new vocabulary which is able to express anything its users might desire. The problem with this is that such an effort would be very expensive, both in terms of time and effort on the part of those creating it, and to the potential users, who have to learn to navigate around it, recognise the new terms, and who have to be supported in using the new terms correctly (or, more often, incorrectly).

The alternative approach to the problem is to evade it, and this is the approach taken in this document. Rather than deprecating the existence of multiple overlapping vocabularies, we embrace it, help interest groups formalise as many of them as are appropriate, and standardise the process of formally declaring the relationships between them. This means that:

- The various vocabularies are allowed to evolve separately, on their own timescales, managed either by the IVOA, individual working groups within the IVOA, or by third parties;

- Specialised vocabularies can be developed and maintained by the community with the most knowledge about a specific topic, ensuring that the vocabulary will have the most appropriate breadth, depth, and precision;

- Users can choose the vocabulary or combination of vocabularies most appropriate to their situation, either when annotating resources, or when querying them; and

- We can retain the previous investments made in vocabularies by users and resource owners.

## 2. SKOS-based vocabularies (informative)

In this section, we introduce the concepts of SKOS-based vocabularies, and the technology of mapping between them. We describe some additional requirements for IVOA vocabularies in the next section, 3. Publishing vocabularies (normative).

### 2.1. Selection of the vocabulary format

After extensive online and face-to-face discussions, the authors have brokered a consensus within the IVOA community that formalised vocabularies should be published at least in SKOS (Simple Knowledge Organization System) format, a W3C draft standard application of RDF to the field of knowledge organisation [std:skosref]. SKOS draws on long experience within the Library and Information Science communities, to address a well-defined set of problems to do with the indexing and retrieval of information and resources; as such, it is a close match to the problem this document is addressing. For more detailed introductory discussion, see the SKOS Primer [isaac08].

ISO 5964 [std:iso5964] defines a number of the relevant terms (ISO 5964:1985=BS 6723:1985; see also [std:bs8723-1] and [std:z39.19]), and some of the (lightweight) theoretical background. The only technical distinction relevant to this document is that between vocabulary and thesaurus: BS-8723-1 defines a controlled vocabulary as a

> prescribed list of terms or headings each one having an assigned meaning [noting that "Controlled vocabularies are designed for use in classifying or indexing documents and for searching them."]

and a thesaurus as a

> Controlled vocabulary in which concepts are represented by preferred terms, formally organised so that paradigmatic relationships between the concepts are made explicit, and the preferred terms are accompanied by lead-in entries for synonyms or quasi-synonyms. (BS-8723-1, sect. 2.39) with a similar definition in ISO-5964 sect. 3.16. The "vocabularies" discussed in this document are therefore more properly termed thesauri, but we will retain the term 'vocabulary' since it is marginally more familiar in the astronomical community.

The paradigmatic relationships in question are those relating a term to a "broader", "narrower" or more generically "related" term. These notions have an operational definition: any resource retrieved as a result of a search on a given term will also be retrievable through a search on that term's "broader term" ("narrower" is a simple inverse, so that for any pair of terms, if A `skos:broader` B, then B `skos:narrower` A; a term may have multiple narrower and broader terms). This is not a subsumption relationship, as there is no implication that the concept referred to by a narrower term is of the same *type* as a broader term. For example, the term "Comet" might have "Nucleus" as a narrower term, but this does not imply that a nucleus is a subclass of comet. Further, the `skos:broader` and `skos:narrower` relationships are not transitive (that is, declaring that A `skos:broader` B and B `skos:broader` C does not imply that A `skos:broader` c). However the SKOS standard includes the notions of `skos:broaderTransitive` and `skos:narrowerTransitive` relations for the subset of vocabularies and systems which would find these useful.

Thus a vocabulary (SKOS or otherwise) is not an ontology. It has lighter and looser semantics than an ontology, and is specialised for the restricted case of resource retrieval. Those interested in ontological analyses can easily transfer the vocabulary relationship information from SKOS to a formal ontological format such as OWL [std:owl].

The purpose of a thesaurus is to help users find resources they might be interested in, be they library books, image archives, or VOEvent packets.

### 2.2. Content and format of a SKOS vocabulary

A published vocabulary in SKOS format consists of a set of "concepts" – an example concept capturing the vocabulary information about spiral galaxies is provided in the Figure below, with the RDF shown in both RDF/XML [std:rdfxml] and Turtle notation [std:turtle] (Turtle is similar to the more informal *Notation3*). The elements of a concept are detailed below.

Figure: examples of a SKOS vocabulary Concept (the `skos:` namespace is presumed to have been declared)

| XML Syntax | Turtle Syntax |
|---|---|

```
<skos:Concept rdf:about="urn:example#spiralGalaxy">
  <skos:altLabel xml:lang="en">spiral nebula</skos:altLabel>
  <skos:broader rdf:resource="urn:example#galaxy"/>
  <skos:definition xml:lang="en"> A spiral galaxy is defined here
    following the definition in ...</skos:definition>
  <skos:hiddenLabel xml:lang="en">spiral glaxy</skos:hiddenLabel>
  <skos:narrower rdf:resource="urn:example#barredSpiralGalaxy"/>
  <skos:notation rdf:datatype="urn:example#_notation">5.1.1</skos:notation>
  <skos:prefLabel xml:lang="de">Spiralgalaxie</skos:prefLabel>
  <skos:prefLabel xml:lang="en">spiral galaxy</skos:prefLabel>
  <skos:related rdf:resource="urn:example#spiralArm"/>
  <skos:scopeNote xml:lang="en">The Sa/Sc/Sd subtypes of Spiral
    galaxies are not represented here,
    and should be noted in image comments.</skos:scopeNote>
</skos:Concept>
<rdf:Description rdf:about="urn:example#_notation">
  <dc:description>Notation described in foo.pdf</dc:description>
</rdf:Description>
```

```
<#spiralGalaxy> a skos:Concept;
  skos:prefLabel
    "spiral galaxy"@en,
    "Spiralgalaxie"@de;
  skos:notation "5.1.1"^^<#_notation>;
  skos:altLabel "spiral nebula"@en;
  skos:hiddenLabel "spiral glaxy"@en;
  skos:definition """ A spiral galaxy is de
    following the definition in ..."""@en;
  skos:scopeNote """The Sa/Sc/Sd subtypes o
    galaxies are not represented here,
    and should be noted in image comments."""
  skos:narrower <#barredSpiralGalaxy>;
  skos:broader <#galaxy>;
  skos:related <#spiralArm> .

# ...
<#_notation> dc:description
  "Notation described in foo.pdf".
```

A SKOS vocabulary is essentially a list of SKOS 'concepts', plus some metadata. Each SKOS concept has some or all of the following features, of which only the first two are required.

- A single URI representing the concept, mainly for use by computers (that is, it is not required to be human-readable). This is a syntactic requirement.

- A single prefered label in each supported language of the vocabulary, for use by humans. This is required.

- Alternative labels which applications may encounter, whether simple synonyms or commonly-used aliases such as "GRB" for "gamma-ray burst", or "Spiral nebula" for "spiral galaxies".

- Hidden labels which capture terms which are sometimes used for the corresponding concept, but which are deprecated in some sense. Very common mis-spellings might arguably be included here, but there is no clear best practice.

- A "notation" code (such as `5.1.1` for spiral galaxies within the AVM), used to uniquely identify a concept within a given scheme. If this is used, then a you should indicate which notation is being used by indicating a 'datatype' for the notation, as discussed in Section 6 of *[std:skosref]*. This is less onerous than it sounds: although this may be defined as a full XML Schema datatype, as illustrated here it need be nothing more sophisticated than an otherwise unused fragment identifier elsewhere in the document, which indicates the notation unambiguously. For our purposes, this serves mostly as documentation, though precision here means that the information is available to a vocabulary processor if it is able to make use of it.

- A definition for the concept, where one exists in the original vocabulary, to give a meaning for the term. This need not be extensive, but it should indicate to someone using the vocabulary what the Concept is intended to refer to, and how precise it is expected to be.

- A scope note to further clarify a definition, or the usage of the concept. While the definition explains the meaning of the term in some abstract sense, the scope note provides practical usage hints to a user of the vocabulary. For example, a scope note might say 'This is not to be confused with...' or 'In the case X, use term Y'.

- A concept may also be involved in any number of relationships with other concepts. The types of relationships are
  - Narrower or more specific concepts, for example a link to the concept representing a "barred spiral galaxy".
  - Broader or more general concepts, for example a link to the token representing galaxies in general.
  - Related concepts, for example a link to the token representing spiral arms of galaxies (note this relationship does not say that spiral galaxies have spiral arms – that would be ontological information of a higher order which is beyond the requirements for information stored in a vocabulary).

Only the URI and preferred label are required; all of the remaining features are optional (see also 3.1. Requirements). Although including these features is desirable in a newly-developed vocabulary, if one is converting an existing vocabulary to SKOS form, it may not be feasible or desirable to add missing information. For example, the vocabularies described in section 4. SKOS versions of existing vocabularies (normative) were derived from existing term lists which had almost none of this value-adding apparatus.

In addition to the information about a single concept, a vocabulary can contain information to help users navigate its structure and contents:

- The "top concepts" of the vocabulary, i.e. those that occur at the top of the vocabulary hierarchy defined by the broader/narrower relationships, can be explicitly stated to make it easier to navigate the vocabulary.

- Concepts that form a natural group can be defined as being members of a "collection".

- Versioning information can be added using change notes.

- Additional metadata about the vocabulary, for example indicating the publisher, must be documented using the Dublin Core metadata set *[std:dublincore]*, *[isaac08]*. At a minimum, the vocabulary's `skos:ConceptScheme` should be annotated with DC Terms "title", "creator", "description" and "created".

- The SKOS standard describes a number of "documentation properties"; these should be used to document provenance of and changes to vocabulary terms.

- A set of mappings between the concepts in different vocabularies see section 2.3. Mapping relationships between vocabularies

Note, the set of mappings between vocabularies has the potential to be circular or create inconsistencies, though this is probably reasonably unlikely in fact. This is in principle out of the control of the vocabulary authors, since vocabularies do not contain mappings, and so this can only be detected dynamically by applications which use the vocabularies.

The DC Terms namespace is `http://purl.org/dc/terms/`, as opposed to the older, and now somewhat deprecated, DC Elements namespace `http://purl.org/dc/elements/1.1/` *[std:dublincore]*, *[isaac08]*. Also, note that the value of the DC Terms `dc:creator` property is an object, not a literal; and the value of the `dc:created` property is a literal string conforming to the W3C profile of ISO 8601 *[std:w3cdtf]*. This standard does not prescribe the content of the (object) value of the `dc:creator` property, but properties from the FOAF namespace are good choices.

Thus a suitable `skos:ConceptScheme` declaration might be:

Figure: examples of a SKOS vocabulary ConceptScheme

**XML Syntax**

```
<rdf:RDF
  xmlns:dc="http://purl.org/dc/terms/"
  xmlns:foaf="http://xmlns.com/foaf/0.1/"
  xmlns:rdf="http://www.w3.org/1999/02/22-rdf-syntax-ns#"
  xmlns:skos="http://www.w3.org/2004/02/skos/core#"
  xml:base="urn:example#">
<skos:ConceptScheme rdf:about="">
  <dc:created>2008-05-08</dc:created>
  <dc:creator>
    <rdf:Description>
      <foaf:name>The IVOA Semantics Working Group</foaf:name>
    </rdf:Description>
  </dc:creator>
  <dc:description xml:lang="en">
    This is a list of keywords ...
  </dc:description>
  <dc:title xml:lang="en">Vocabulary for ...</dc:title>
</skos:ConceptScheme>
<!-- ... -->
</rdf:RDF>
```

**Turtle Syntax**

```
@prefix dc: <http://purl.org/dc/terms/> .
@prefix foaf: <http://xmlns.com/foaf/0.1/> .
@prefix skos: <http://www.w3.org/2004/02/skos/core#> .

<> a skos:ConceptScheme ;
  dc:created "2008-05-08" ;
  dc:title "Vocabulary for ..."@en ;
  dc:creator [
    foaf:name
      "The IVOA Semantics Working Group"
  ];
  dc:description
    "This is a list of keywords ..."@en.
  # etc
```

## 2.3. Mapping relationships between vocabularies

There already exist several vocabularies in the domain of astronomy. Instead of attempting to replace all these existing vocabularies, which have been developed to achieve different aims and user groups, we embrace them. This requires a mechanism to relate the concepts in the different vocabularies.

Part of the SKOS working draft standard *[std:skosref]* allows a concept in one vocabulary to be related to a concept in another vocabulary.

There are four types of relationship provided to capture the relationships between concepts in vocabularies, which are similar to those defined for relationships between concepts within a single vocabulary. The types of mapping relationships are as follows.

- Equivalence between concepts, i.e. the concepts in the different vocabularies refer to the same real world entity. This is captured with the RDF statement

    ```
    AAkeys:#Cosmology skos:exactMatch avm:#Cosmology
    ```

    which states that the cosmology concept in the A&A Keywords is the same as the cosmology concept in the AVM. (Note the use of an external namespaces `AAkeys` and `avm` which must be defined within the document.)

- Broader concept, i.e. there is not an equivalent concept but there is a more general one. This is captured with the RDF statement

    ```
    AAkeys:#Moon skos:broadMatch avm:PlanetSatellite
    ```

    which states that the AVM concept "Planet Satellite" is a more general term than the A&A Keywords concept "Moon".

- Narrower concept, i.e. there is not an equivalent concept but there is a more specific one. This is captured with the RDF statement

    ```
    AAkeys:#IsmClouds skos:narrowMatch avm:#NebulaAppearanceDarkMolecularCloud
    ```

    which states that the AVM concept "Nebula Appearance Dark Molecular Cloud" is more specific than the A&A Keywords concept "ISM Clouds".

- Related concept, i.e. there is some form of associative relationship. This is captured with the RDF statement

    ```
    AAkeys:#BlackHolePhysics skos:relatedMatch avm:#StarEvolutionaryStageBlackHole
    ```

    which states that the A&A Keywords concept "Black Hole Physics" has an association with the AVM concept "Star Evolutionary Stage Black Hole".

The semantic mapping relationships have certain properties. The broadMatch relationship has the narrowMatch relationship as its inverse and the exactMatch and relatedMatch relationships are symmetrical. The consequence of these properties is that if you have a mapping from concept A in one vocabulary to concept B in another vocabulary then you can infer a mapping from concept B to concept A.

## 2.4. Vocabulary versions

The document *[kendall08]* discusses good practice for managing RDF vocabularies. At the time of writing (2008 May) this is still an editor's draft, and it itself notes that good practice in this area is not yet fully stable, so our recommendations here are necessarily tentative, and in some places restricted to the relatively small vocabularies (100s to 1000s of terms) we expect to encounter in the VO. We expect to adjust or enhance this advice in future editions of this Recommendation, as best practice evolves, or as we gain more experience with the relevant vocabularies.

We must distinguish between versions of a vocabulary, and versions of the description of a vocabulary. In the former case, we are concerned with the presence or absence of certain concepts, such as "star" or "GRB", and expect that there will be some reasonably stable relationship between the concept URI and the real-world concept it refers to. In the latter case, we are concerned with the technicalities of associating a concept URI with its labels, its description, and with other related concept URIs. While it is true that there are epistemological commitments involved in the simple act of naming (and the terms "GRB" and "planet" remind us that there is knowledge implicit within a name), it is the latter case that generally represents the *knowledge* we have of an object, and it is this knowledge which we must version. For example, the concept of "planet" is a stable one, and so should not be versioned, but the *definition* of a planet was changed by the IAU in 2006, so that the description of a vocabulary term such as "planet" would have changed then, and should be versioned.

In consequence, *the concept URIs should not carry version information*. The partial exception to this is when a vocabulary undergoes a major restructuring, as a result of the terms in it becoming significantly incoherent – for example, we might imagine the IAU93 thesaurus being updated to form an IAU 200x thesaurus – but in this case we should regard the result as a new vocabulary, rather than simply an adjusted version of an old one.

All the terms in the SKOS vocabulary appear in an unversioned namespace, and once in the vocabulary they are not removed *[kendall08]*. Successive versions of the vocabulary description describe the vocabulary terms as "unstable", "testing", "stable" or "deprecated".

The Dublin Core namespaces are managed in a similar way *[dc:namespaces]*. The namespace URIs, which act as common prefixes to the DC terms, and which are defined using a "hash URI" strategy, in RDF terms, have no version numbers, so that the namespace for the DC terms vocabulary is `http://purl.org/dc/terms/`. Terms such as `http://purl.org/dc/terms/extent` then 302-redirect to a URL which, for administrative convenience, happens to contain a release date, but which resolves to RDF which defines the unversioned term `http://purl.org/dc/terms/extent`. This file includes the following content (translated into Turtle from the original RDF/XML for legibility).

```
@prefix rdf: <http://www.w3.org/1999/02/22-rdf-syntax-ns#> .
@prefix skos: <http://www.w3.org/2004/02/skos/core#> .
@prefix dcam: <http://purl.org/dc/dcam/> .
@prefix dcterms: <http://purl.org/dc/terms/> .
@prefix rdfs: <http://www.w3.org/2000/01/rdf-schema#> .

<http://purl.org/dc/terms/>
  dcterms:title """DCMI Namespace for metadata terms
            in the http://purl.org/dc/terms/ namespace"""@en-us;
  rdfs:comment """To comment on this schema,
            please contact dcmifb@dublincore.org."""@en-us;
  dcterms:publisher "The Dublin Core Metadata Initiative"@en-us;
  dcterms:modified "2008-01-14" .

dcterms:extent
  rdfs:label "Extent"@en-us;
  rdfs:comment "The size or duration of the resource."@en-us;
  rdfs:isDefinedBy <http://purl.org/dc/terms/>;
  dcterms:issued "2000-07-11";
  dcterms:modified "2008-01-14";
  a rdf:Property;
  dcterms:hasVersion <http://dublincore.org/usage/terms/history/#extent-003>;
  rdfs:range dcterms:SizeOrDuration;
  rdfs:subPropertyOf <http://purl.org/dc/elements/1.1/format>,
    dcterms:format .
...
```

This includes the definition of the (unversioned) `http://purl.org/dc/terms/extent` concept, along with semantic knowledge about the concept

(`rdfs:subPropertyOf`) as of 2008-01-14, plus other editorial (`dcterms:modified`) and definitional (`rdfs:isDefinedBy`) metadata.

# 3. Publishing vocabularies (normative)

## 3.1. Requirements

A vocabulary which conforms to this IVOA standard has the following features.

### 3.1.1. Dereferenceable namespace

The namespace of the vocabulary MUST be dereferenceable on the web. That is, typing the namespace URL into a web browser will produce human-readable documentation about the vocabulary. In addition, the namespace URL SHOULD return an RDF version of the vocabulary if it is retrieved with one of the RDF MIME types in the HTTP Accept header. At the time of writing, the only fully standardised RDF MIME type is `application/rdf+xml` for RDF/XML, but `text/rdf+n3` and `text/turtle` are the proposed types for Notation3 [_notation3_] and Turtle [_std:turtle_], respectively.

_Rationale: These prescriptions are intended to be compatible with the patterns described in [_berrueta08_] and [_sauermann08_], and vocabulary distributors_ SHOULD _follow these patterns where possible._

### 3.1.2. Long-term availability

The files defining a vocabulary, including those of superseded versions, SHOULD remain permanently available. There is no requirement that the namespace URL be at any particular location, although the IVOA web pages, or a journal publisher's web pages, would likely be suitable archival locations.

### 3.1.3. Distribution format

Vocabularies MUST be made available for distribution as SKOS RDF files in RDF/XML [_std:rdfxml_] format. A human readable version in Turtle [_std:turtle_] format SHOULD also be made available. As an alternative to Turtle, vocabularies may be made available in that subset of Notation3 [_notation3_] which is compatible with Turtle; if Turtle or Notation3 is being served, it is prudent to support both `text/rdf+n3` and `text/turtle` as MIME types in the `Accept` header of the HTTP request.

A publisher MAY make available RDF in other formats, or other supporting files. A publisher MUST make available at least some human-readable documentation – see section 3.3. Good practices when serving vocabularies on the web for a discussion of the mechanics here.

_Rationale: this does imply that the vocabulary source files can only realistically be parsed using an RDF parser. An alternative is to require that vocabularies be distributed using a subset of RDF/XML which can also be naively handled as traditional XML; however as well as creating an extra standardisation requirement, this would make it effectively infeasible to write out the distribution version of the vocabulary using an RDF or general SKOS tool._

### 3.1.4. Clearly versioned vocabulary

The vocabulary _namespace_ SHOULD NOT be versioned, but it SHOULD be easy to retrieve earlier versions of the RDF describing the vocabulary. See the discussion in section 2.4. Vocabulary versions for the rationale for this, and see section 3.3. Good practices when serving vocabularies on the web for a discussion of its implications for the way that vocabularies are served on the web.

### 3.1.5. Concepts must have labels

Each concept MUST have both a URI naming it, and a human-readable `skos:prefLabel` (the first is a syntactic requirement of SKOS, and so is trivially satisfied; the second is an extension to SKOS).

### 3.1.6. No restrictions on source files

This Recommendation does not place any restrictions on the format of the files managed by the maintenance process, as long as the distributed files are as specified above.

## 3.2. Good practices of vocabulary design

This standard imposes a number of requirements on conformant vocabularies (see 3.1. Requirements). In this section we list a number of good practices that IVOA vocabularies SHOULD abide by. Some of the prescriptions below are more specific than good-practice guidelines for vocabularies in general.

The adoption of the following guidelines will make it easier to use vocabularies in generic VO applications. However, VO applications MUST be able to accept any vocabulary that complies with the latest SKOS standard [_std:skosref_] (this is a syntactical requirement, and does not imply that an application will necessarily understand the terms in an alien vocabulary, although the presence of mappings to a known vocabulary should allow it to derive some benefit).

1. SKOS concepts are identified by strings which are, syntactically, the fragment part of a URI. As such, these identifiers must be composed of characters from a restricted set. The URI specification [_std:rfc3986_] permits a broad range of characters in a URI fragment, but for simplicity, we recommend here that concepts be identified by more restricted strings, and SHOULD match the regular expression
   `^[A-Za-z0-9][A-Za-z0-9._-]*$`.

2. For the sake of expert users working directly with the concept URIs, the concept identifiers SHOULD be kept in human-readable form, directly reflecting the concept's label, or place in a hierarchy, and not be semi-random identifiers only (for example, use `spiralGalaxy`, not `t1234567`). [Rationale: When working with very large or complicated vocabularies and ontologies, it is useful to have opaque concept labels, to avoid confusion arising from the intuitive semantics of a recognisable label. This is less of a danger with the simpler vocabularies we are discussing here, so we can safely retain the convenience of recognisable concept identifiers.]

3. When developing a _new_ vocabulary, standard thesaurus practice indicates that english language labels for concrete concepts should be pluralised, though abstract concepts remain singular – thus `"galaxies"@en`, but `"astronomy"@en`. Thesaurus practice in other languages uses the singular for all cases.

4. However, when adapting an existing vocabulary into the SKOS format, implementors SHOULD NOT change the labels (for example changing the grammatical number) beyond any changes necessarily required by SKOS.

5. Each concept SHOULD have a definition (`skos:definition`) that constitutes a short description of the concept which could be adopted by an application using the vocabulary. Each concept SHOULD have additional documentation using SKOS Notes or Dublin Core terms as appropriate (see [*std:skosref*]). In practice, this requirement is rather difficult to satisfy when converting pre-existing structured vocabularies to SKOS, since these frequently provide only labels, without fuller descriptions or scope notes; in this case, vocabulary developers SHOULD NOT add trivial descriptions which do little more than echo the label text.

6. The language localisation SHOULD be declared where appropriate, in preferred labels, alternate labels, definitions, and the like. Thus, use `"galaxies"@en`, rather than just `"galaxies"`, even if there is only one language being used in the vocabulary. According to the SKOS reference, the list of language tags is that of [*std:bcp47*], which describes a register of two-character language tags which is effectively identical with that of [*std:iso639-1*].

7. Relationships ("broader", "narrower", "related") between concepts SHOULD be present, but are not required; if used, they SHOULD be complete (thus all "broader" links have corresponding "narrower" links in the referenced entries and "related" entries link each other).

8. A vocabulary SHOULD indicate the "top concepts" in the vocabulary (namely those concepts that do not have any "broader" relationships), using the `skos:hasTopConcept` relation. This should be done only if the vocabulary is structured enough that this is useful. [Rationale: "top concepts" can be of use as initial hints for a user interface which aims to help a user navigate through a vocabulary starting from "the top"; if a vocabulary is so flat that these "top concepts" would be too numerous, then the vocabulary developer may reasonably decide not to add them]

9. A vocabulary MUST contain only one `skos:ConceptScheme`.

10. The SKOS standard describes some good practices for vocabulary maintenance, such as using `<skos:changeNote>` and the like, and these are elaborated in the (currently draft) note [*kendall08*]. At a minimum, the vocabulary's `skos:ConceptScheme` MUST be annotated with properties in the DC Terms namespace `http://purl.org/dc/terms/`, namely "dc:title", "dc:creator", "dc:created" and "dc:description", with values of an appropriate type, as illustrated above in 2.2. Content and format of a SKOS vocabulary. Publishers SHOULD respect such good practices as are available to direct vocabulary development and maintenance.

11. Publishers SHOULD publish "mappings" between their vocabularies and other commonly used vocabularies (see 5. Mapping vocabularies (informative)). These MUST be external to the defining vocabulary document so that the vocabulary can be used independently of the publisher's mappings. The mapping file MUST be annotated with DC terms "title", "creator", "created" and "description" [*std:dublincore*], [*isaac08*].

## 3.3. Good practices when serving vocabularies on the web

The W3C Interest Group Note *Cool URIs for the Semantic Web* [*sauermann08*] presents guidelines for the effective use of URIs when serving web documents and concepts on the Semantic Web. When providing vocabularies to the VO, we RECOMMEND that publishers conform to these guidelines in general. We make some further observations below.

The "Cool URIs" guidelines describe a number of desirable features in this context, namely simplicity, stability and manageability. Section 4.5 of the document describes these features as follows (quoted directly).

**Simplicity**
Short, mnemonic URIs will not break as easily when sent in emails and are in general easier to remember, e.g. when debugging your Semantic Web server.
**Stability**
Once you set up a URI to identify a certain resource, it should remain this way as long as possible. Think about the next ten years. Maybe twenty. Keep implementation-specific bits and pieces such as .php and .asp out of your URIs, you may want to change technologies later.
**Manageability**
Issue your URIs in a way that you can manage. One good practice is to include the current year in the URI path, so that you can change the URI-schema each year without breaking older URIs. Keeping all 303 URIs on a dedicated subdomain, e.g. `http://id.example.com /alice`, eases later migration of the URI-handling subsystem.

We endorse this advice in this Recommendation: VO vocabularies SHOULD use URIs which have these properties. The advice in the third point is a general point about maintaining the general URI namespace on a particular server, and is not about versioning vocabulary namespaces.

The "Cool URIs" document also describes two broad strategies for making these URIs available on the web, which they name *303 URIs* and *hash URIs* (see the document, section 4, for descriptions). They note that the *hash URI* strategy "should be preferred for rather small and stable sets of resources that evolve together. The ideal case[s] are RDF Schema vocabularies and OWL ontologies, where the terms are often used together, and the number of terms is unlikely to grow out of control in the future." Since this is the case for the (relatively small) SKOS vocabularies this Recommendation discusses, and since an application will generally want to use the complete vocabulary rather than only single concepts, we suggest that vocabularies conformant to this Recommendation SHOULD be distributed as *hash URI* ones.

Common to the two strategies above is the insistence that the vocabulary URIs *are HTTP URIs which are retrievable on the web* – they differ only in the practicalities of achieving this. The strategies also share the expectation that the vocabulary URIs are retrievable both as RDF (machine-readable) and as HTML (providing documentation for humans). We elevate this to a requirement of this Recommendation: vocabulary terms MUST be HTTP URIs which MUST be dereferenceable as both RDF and HTML using the mechanism appropriate to the URI naming strategy. Specifically, dereferencing the vocabulary namespace MUST result in a 303 redirection to a URL which produces a content type appropriate to the content type in any `Accept` header in the initial request.

## 3.4. Example and recipe: serving the A&A vocabulary (informative)

While [*sauermann08*], as quoted in section 3.3. Good practices when serving vocabularies on the web, discusses the design of the URIs naming concepts, it says little about the mechanics of making these available on the web. We refer vocabulary publishers to the recipe advice contained in [*berrueta08*], which we illustrate here with an explicit recipe, in the case of the *hash URI* strategy.

The A&A vocabulary has the namespace `http://www.ivoa.net/rdf/Vocabularies/AAkeys` (see 4.1. The Astronomy & Astrophysics Keyword List). In accordance with the above guidelines, this namespace URI is dereferenceable, and if you enter the URI into a web browser, you will end up at a page describing the vocabulary. The way this works can be illustrated by using the `curl` utility to dereference the URI (URIs are cropped for legibility):

```
% curl http://[...]/rdf/AAkeys
HTTP/1.1 303 See Other
Date: Thu, 08 May 2008 14:07:12 GMT
Server: Apache
Location: http://[...]/rdf/vocabularies-20091007/AAkeys/AAkeys.html
Connection: close
Content-Type: text/html; charset=utf-8

<title>Redirected</title>
<p>See <a href='http://[...]/rdf/vocabularies-20091007/AAkeys/AAkeys.html'
```

```
>elsewhere</a></p>
```

The server has responded to the HTTP GET for the URI with a 303 response, and a `Location` header, pointing to the HTML representation of this thing. In this example, the server has included a brief HTML explanation in case a human happens to see this response.

If we instead request an RDF representation, by stating a desired MIME type in the HTTP `Accept` header, we get a slightly different response:

```
% curl --head -H accept:text/turtle http://[...]/rdf/AAkeys
HTTP/1.1 303 See Other
Date: Thu, 08 May 2008 14:11:28 GMT
Server: Apache
Location: http://[...]/rdf/vocabularies-20091007/AAkeys/AAkeys.ttl
Connection: close
Content-Type: text/html; charset=iso-8859-1
```

This is also a 303 response, but the `Location` header this time points to an RDF file in Turtle syntax, which we can now retrieve normally.

```
% curl --include http://[...]/rdf/vocabularies-20091007/AAkeys/AAkeys.ttl
HTTP/1.1 200 OK
Date: Thu, 08 May 2008 14:13:35 GMT
Server: Apache
Content-Type: text/turtle; charset=utf-8

@base <http://[...]/rdf/AAkeys> .
@prefix rdf: <http://www.w3.org/1999/02/22-rdf-syntax-ns#> .
@prefix dc: <http://purl.org/dc/terms/> .
@prefix rdfs: <http://www.w3.org/2000/01/rdf-schema#> .
@prefix foaf: <http://xmlns.com/foaf/0.1/> .
@prefix : <http://www.w3.org/2004/02/skos/core#> .

<>
  dc:created "2008-05-08" ;
  dc:title "Vocabulary for Astronomy & Astrophysics Journal keywords (Version wd-1.0)"@en ;
  dc:creator [
    foaf:name "The IVOA Semantics Working Group"
  ];
  dc:description "This is a list of journal keywords developed by ...";
  a :ConceptScheme ;
# and so on...
```

Note that the base URI in the returned RDF still refers to the unversioned concept names.

This behaviour is controlled by (in this case) an Apache `.htaccess` file which looks like this:

```
AddType application/rdf+xml .rdf
# The MIME type for .n3 should be text/rdf+n3, not application/n3:
# see MIME notes at <http://www.w3.org/2000/10/swap/doc/changes.html>
#
# The MIME type for Turtle is text/turtle, though this has not
# completed its registration: see
# <http://www.w3.org/TeamSubmission/turtle/#sec-mediaReg>
AddType text/rdf+n3 .n3
AddType text/turtle .ttl
# For Charset types, see <http://www.iana.org/assignments/character-sets>
AddCharset UTF-8 .n3
AddCharset UTF-8 .ttl

RewriteEngine On
# This will match the directory where this file is located.
RewriteBase /users/norman/ivoa/vocabularies/rdf

RewriteCond %{HTTP_ACCEPT} application/rdf\+xml
RewriteRule ^(AAkeys|AVM|UCD|IVOAT|IAUT93)$ vocabularies-20091007/$1/$1.rdf [R=303]

RewriteCond %{HTTP_ACCEPT} text/rdf\+n3 [OR]
RewriteCond %{HTTP_ACCEPT} application/n3 [OR]
RewriteCond %{HTTP_ACCEPT} text/turtle
RewriteRule ^(AAkeys|AVM|UCD|IVOAT|IAUT93)$ vocabularies-20091007/$1/$1.ttl [R=303]

# Any other accept header, including none: make the .html version the default
RewriteRule ^(AAkeys|AVM|UCD|IVOAT|IAUT93)$ vocabularies-20091007/$1/$1.html [R=303]
```

These various `RewriteRule` statements examine the content of the HTTP `Accept` header, and return 303-redirections to the appropriate actual resource.

Note that the namespace remains unversioned throughout the maintenance history of this vocabulary, even though the actual RDF files being returned might change as labels or relationships are adjusted. Previous versions of the vocabulary RDF will remain available, though they will no longer be served by dereferencing the namespace URL.

## 4. SKOS versions of existing vocabularies (normative)

The intent of having the IVOA adopt SKOS as the preferred format for astronomical vocabularies is to encourage the creation and management of diverse vocabularies by competent astronomical groups, so that users of the VO and related resources can benefit directly and dynamically without the intervention of the IAU or IVOA. In this section, we include SKOS versions of several vocabularies which already exist, and have some usage within astronomy. This illustrates the use of SKOS with real vocabularies, but also, more importantly, makes these vocabularies available for immediate use in VO applications.

The vocabularies described below are included, as SKOS files, in the distributed version of this standard. These vocabularies have stable URLs, and may be cited and used indefinitely. These vocabularies will not, however, be developed as part of the maintenance of this standard. Interested groups, within and outwith the IVOA, are encouraged to take these as a starting point and absorb them within existing processes.

The exceptions to this rule are the constellation vocabulary, provided here mainly for didactic purposes, and the proposed IVOA Thesaurus, which is being developed as a separate project and whose aim is to provide a corrected, more user-friendly, more complete, and updated version of the 1993 IAU thesaurus. Although work on the IVOA Thesaurus is on-going, the fact that it is largely based on the IAU thesaurus means that it is already a very useful resource, so a usable snapshot of this vocabulary will be published with the other examples.

We provide a set of SKOS files representing the vocabularies which have been developed, and an example mapping file between the A&A keywords and the AVM Taxonomy. These vocabularies have base URIs starting `http://www.ivoa.net/rdf/Vocabularies`, and can be downloaded at the URL

    `http://www.ivoa.net/rdf/Vocabularies/vocabularies-20091007.tar.gz`

## 4.1. The Astronomy & Astrophysics Keyword List

Namespace: `http://www.ivoa.net/rdf/Vocabularies/AAkeys`

This vocabulary is based on a set of keywords maintained jointly by the publishers of the journals *Astronomy and Astrophysics* (A&A), *Monthly Notices of the Royal Astronomical Society* (MNRAS) and the *Astrophysical Journal* (ApJ), and updated on an annual basis. As noted in the introduction, an analysis of these keywords *[preitemartinez07]* indicates that the different journals are slightly inconsistent with each other; we have rather arbitrarily used the 2008 list from the A&A web site. The intended usage of the vocabulary is to tag articles with descriptive keywords to aid searching for articles on a particular topic.

The keywords are organised into categories which have been modelled as hierarchical relationships. Additionally, some of the keywords are grouped into collections which has been mirrored in the SKOS version. The vocabulary contains no definitions or related links as these are not provided in the original keyword list, and only a handful of alternative labels and scope notes that are present in the original keyword list.

## 4.2. The AVM Taxonomy

Namespace: `http://www.ivoa.net/rdf/Vocabularies/AVM`.

This vocabulary is published by the *Virtual Astronomy Multimedia Project* (`http://www.virtualastronomy.org/` and *[std:avm]*) to allow images to be tagged with keywords that are relevant for the public. It consists of a set of keywords organised into an enumerated hierarchical structure. Each term consists of a taxonomic number and a label. There are no definitions, scope notes, or cross references.

When converting the AVM into SKOS, The IVOA working group decided to model the taxonomic number as both a `skos:notation` and an alternative label. Since some terms are duplicated, the token for a term consists of the taxonomical number to avoid ambiguity and to keep the tokens short.

## 4.3. The UCD1+ Vocabulary

Namespace: `http://www.ivoa.net/rdf/Vocabularies/UCD`.

The UCD standard is an officially sanctioned and managed vocabulary of the IVOA *[std:ucd]*. The normative document is a simple text file containing entries consisting of tokens (for example `em.IR`), a short description, and usage information ("syntax codes" which permit UCD tokens to be concatenated). The form of the tokens implies a natural hierarchy: `em.IR.8-15um` is obviously a narrower term than `em.IR`, which in turn is narrower than `em`.

Given the structure of the UCD1+ vocabulary, the natural translation to SKOS consists of preferred labels equal to the original tokens (the UCD1 words include dashes and periods), vocabulary tokens created using guidelines in 3.2. Good practices of vocabulary design (for example, "emIR815UM" for `em.IR.8-15um`), direct use of the definitions, and the syntax codes placed in usage documentation: `<skos:scopeNote>UCD syntax code: P</skos:scopeNote>`

Note that the SKOS document containing the UCD1+ vocabulary does NOT consistute the official version: the normative document is still the text list. However, on the long term, the IVOA may decide to make the SKOS version normative, since the SKOS version contains all of the information contained in the original text document but has the advantage of being in a standard format easily read and used by any application on the semantic web whilst still being usable in the current ways.

## 4.4. The 1993 IAU Thesaurus

Namespace: `http://www.ivoa.net/rdf/Vocabularies/IAUT93`.

The IAU Thesaurus consists of concepts with mostly capitalised labels and a rich set of thesaurus relationships ("BT" for "broader term", "NT" for "narrower term", and "RT" for "related term"). The thesaurus also contains "U" (for "use") and "UF" ("use for") relationships. In a SKOS model of a vocabulary these are captured as alternative labels. A separate document contains translations of the vocabulary terms in five languages: English, French, German, Italian, and Spanish. Enumerable concepts are plural (for example "SPIRAL GALAXIES") and non-enumerable concepts are singular (for example "STABILITY"). Finally, there are some usage hints like "combine with other", which have been modelled as scope notes.

In converting the IAU Thesaurus to SKOS, we have been as faithful as possible to the original format of the thesaurus. Thus, preferred labels have been kept in their uppercase format.

The IAU Thesaurus has been unmaintained since its initial production in 1993; it is therefore significantly out of date in places. This vocabulary is published for the sake of completeness, and to make the link between the evolving vocabulary work and any uses of the 1993 vocabulary which come to light. We do not expect to make any future maintenance changes to this vocabulary, and would expect the IVOAT vocabulary, based on this one, to be used instead (see 4.5. Towards an IVOA Thesaurus (informative)).

## 4.5. Towards an IVOA Thesaurus (informative)

While it is true that the adoption of SKOS will make it easy to publish and access different astronomical vocabularies, the fact is that there is no vocabulary which makes it easy to jump-start the use of vocabularies in generic astrophysical VO applications: each of the previously developed vocabularies has their own limits and biases. For example, the IAU Thesaurus provides a large number of entries, copious relationships, and translations to four other languages, but there are no definitions, many concepts are now only useful for historical purposes (for example many photographic or historical instrument entries), some of the relationships are false or outdated, and many important or newer concepts and their common abbreviations are missing.

Despite its faults, the IAU Thesaurus constitutes a very extensive vocabulary which could easily serve as the basis vocabulary once we have removed its most egregious faults and extended it to cover the most obvious semantic holes. To this end, a heavily revised IAU thesaurus is in preparation for use within the IVOA and other astronomical contexts. The goal is to provide a general vocabulary foundation to which other, more specialised, vocabularies can be added as needed, and to provide a good "lingua franca" for the creation of vocabulary mappings.

## 4.6. A Constellation Name Vocabulary (informative)

This vocabulary is presented as a simple example of an astronomical vocabulary for a very particular purpose, such as handling constellation information like that commonly encountered in variable star research. For example, "SS Cygni" is a cataclysmic variable located in the constellation "Cygnus". The name of the star uses the genitive form "Cygni", but the alternate label "SS Cyg" uses the standard abbreviation "Cyg". Given the constellation vocabulary, all of these forms are recorded together in a computer-manipulatable format. Various incorrect forms should probably be represented in SKOS hidden labels.

The `<skos:ConceptScheme>` contains a single `<skos:TopConcept>`, "constellation"

| XML Syntax | Turtle Syntax |
|---|---|

```
<rdf:RDF
    xmlns:dc="http://purl.org/dc/terms/"
    xmlns:foaf="http://xmlns.com/foaf/0.1/"
    xmlns:rdf="http://www.w3.org/1999/02/22-rdf-syntax-ns#"
    xmlns="http://www.w3.org/2004/02/skos/core#"
    xml:base="urn:example#">
  <ConceptScheme rdf:about="">
    <dc:created>2009-08-20</dc:created>
    <dc:creator>
      <rdf:Description>
        <foaf:name>Rick Hessman</foaf:name>
      </rdf:Description>
    </dc:creator>
    <dc:description xml:lang="en">
      A sample constellation vocabulary
    </dc:description>
    <dc:title xml:lang="en">Constellations</dc:title>
  </ConceptScheme>
  <Concept rdf:about="urn:example#constellation">
    <definition>IAU-sanctioned constellation names</definition>
    <inScheme rdf:resource=""/>
    <narrower rdf:resource="urn:example#Andromeda"/>
    <narrower rdf:resource="urn:example#Cygnus"/>
    <narrower rdf:resource="urn:example#Vulpecula"/>
    <prefLabel xml:lang="en">constellation</prefLabel>
  </Concept>
  <!-- ... -->
</rdf:RDF>
```

```
<> a :ConceptScheme;
    dc:created "2009-08-20";
    dc:title "Constellations"@en;
    dc:creator [ foaf:name "Rick Hessman" ];
    dc:description "A sample constellation vocabulary"@en

<#constellation> a :Concept;
    :inScheme <>;
    :prefLabel "constellation"@en;
    :definition "IAU-sanctioned constellation names";
    :narrower <#Andromeda>;
    :narrower <#Cygnus>;
    ...
    :narrower <#Vulpecula>.
```

and the entry for "Cygnus" is

```
<Concept rdf:about="urn:example#Cygnus">
    <altLabel>Cyg</altLabel>
    <altLabel>Cygni</altLabel>
    <broader rdf:resource="urn:example#constellation"/>
    <definition xml:lang="en">Cygnus</definition>
    <inScheme rdf:resource=""/>
    <prefLabel>Cygnus</prefLabel>
    <scopeNote>Cygnus is nominative form;
    the alternative labels are the genitive and
    short forms</scopeNote>
</Concept>
```

```
<#Cygnus> a :Concept;
    :inScheme <>;
    :prefLabel "Cygnus";
    :definition "Cygnus"@en;
    :altLabel "Cygni";
    :altLabel "Cyg";
    :broader <#constellation>;
    :scopeNote """Cygnus is nominative form;
    the alternative labels are the genitive and
    short forms"""@en .
```

Note that SKOS alone does not permit the distinct differentiation of genitive forms and abbreviations, but the use of alternate labels is more than adequate enough for processing by VO applications where the difference between "SS Cygni", "SS Cyg", and the incorrect form "SS Cygnus" is probably irrelevant.

We haven't declared languages for the `:prefLabel` or `:altLabel` properties: `@en` doesn't seem quite right, `@la` seems pompous, and the "no linguistic content" tag `@zxx` seems confusing and still inaccurate. In this case, it seems that the *should* element of the relevant best practice in Sect. 3.2. Good practices of vocabulary design should be invoked, and the language tag omitted.

# 5. Mapping vocabularies (informative)

Part of the motivation for formalising vocabularies within the VO is to support *mapping between vocabularies*, so that an application which understands, or can natively process, one vocabulary, can use a mapping to provide at least partial support for data described using another vocabulary. Section 10 of the SKOS standard *[std:skosref]* describes a number of properties for expressing such matches, and we anticipate that we will shortly see explicit mappings between vocabularies, produced either by vocabulary maintainers, describing the relationships between their own vocabularies and others, or by third parties, asserting such relationships as an intellectual contribution of their own.

The vocabularies distributed in association with this document include one non-exhaustive mapping mapping file between A&A keywords and the AVM taxonomy, as an example of how such mappings will appear.

Mapping: http://www.ivoa.net/rdf/Vocabularies/AAkeys2AVMMapping.

---

# Appendices

# A. Change history

## Changes between 1.18 and 1.19/REC

**None**

## Changes between 1.17 and 1.18

- Fixed inline examples so that they abide by the document's prescriptions, and pass the validator.

- The SKOS reference is now (finally) a W3C Recommendation.

## Changes between 1.16 and 1.17

- Update status of SKOS standard (now a W3C PR, so *close* to REC)

- Slightly clarified best practice for vocabulary definition text, and clarified what SKOS features are and are not required ([issues](#) 10 and 12)

- Refer to DC Terms elements, rather than the now-deprecated DC Elements terms.

- Clarified the best practices of Section [3.2. Good practices of vocabulary design](#), and added a validator application. For detailed discussion of the changes required, see `projects/vocabularies/src/code/validator/README`

- Updated the distributed vocabularies to ensure that they are valid according to the validator

- In both the document and the distributed vocabularies, changed the SKOS namespace from `http://www.w3.org/2008/05/skos#` to `http://www.w3.org/2004/02/skos/core#`, reflecting the SKOS WG's decision to make this reversion.

- Change author affiliation for AJGG

- Various typo-level changes, to phraseology, references, URLs, etc

# References

**[berrueta08] Diego Berrueta and Jon Phipps.**
Best practice recipes for publishing RDF vocabularies. W3C Working Group Note, August 2008. [Online].
**[dc:namespaces] Andy Powell and Harry Wagner, editors.**
Namespace policy for the dublin core metadata initiative (DCMI). DCMI Recommendation, July 2007. [Online].
**[hessman05] F V Hessman.**
VOConcepts — a proposed UCD for astronomical objects, events, and processes. [Online, cited January 2008].
**[isaac08] Antoine Isaac and Ed Summers.**
SKOS simple knowledge organization system primer. W3C Working Group Note, August 2009. [Online].
**[kendall08] Elisa Kendall and Vit Novacek, editors.**
Principles of good practice for managing RDF vocabularies and OWL ontologies. W3C Editor's Draft, March 2008. [Online].
**[lortet94] M-C Lortet, S Borde, and F Ochsenbein.**
Second reference dictionary of the nomenclature of celestial objects. *Astron. Ap. Supp*, **107** pp. 193-218, 1994. [Online].
**[lortet94a] M-C Lortet, S Borde, and F Ochsenbein.**
The second reference dictionary of the nomenclature of celestial objects (solar system excluded). volumes i, ii.. Technical Report 24, Centre de Données astronomique de Strasbourg, 1994. [Online].
**[notation3] Tim Berners-Lee.**
Notation 3. Web page, 2006. [Online, cited 2008 May 8].
**[preitemartinez07] Andrea Preite Martinez and Soizick Lesteven.**
Astronomical keywords in the era of the virtual observatory. IVOA Note, IVOA, 2007. [Online].
**[sauermann08] Leo Sauermann and Richard Cyganiak.**
Cool URIs for the semantic web. W3C Interest Group Note, March 2008. [Online].
**[shobbrook93] Robyn M Shobbrook and Robert R Shobbrook.**
*The Astronomy Thesaurus*. Anglo-Australian Observatory, 1993. [Online].
**[std:avm] Robert Hurt, Lars Lindberg Christensen, and Adrienne Gauthier.**
Virtual astronomy metadata project. [Online, cited 2008 June 23].
**[std:bcp47] A Phillips and M Davis.**
Tags for identifying languages. Internet RFC, September 2006. Also known as RFC 4646. [Online].
**[std:bs8723-1] Structured vocabularies for information retrieval — guide — definitions, symbols and abbreviations (BS 8723-1:2005).**
British Standard, 2005.
**[std:dublincore] DCMI Usage Board.**
DCMI metadata terms. DCMI Recommendation, 2006. [Online].
**[std:iso5964] Documentation — guidelines for the establishment and development of multilingual thesauri (ISO 5964:1985=BS 6723:1985).**
International Standard, 1985.
**[std:iso639-1] Codes for the representation of names of languages— part 1: alpha-2 code (ISO 639-1:2002).**
International Standard, 2002. [Online].
**[std:ivoa-astro-onto] Sébastien Derriere, Andrea Preite Martinez, and Alexandre Richard, editors.**
Ontology of astronomical object types. IVOA Note, 2007. [Online].
**[std:owl] World Wide Web Consortium.**
The web ontology language. [Online].
**[std:rdfxml] Dave Beckett.**
RDF/XML syntax specification (revised). W3C Recommendation, February 2004. [Online].
**[std:rfc2119] S Bradner.**
Key words for use in RFCs to indicate requirement levels. RFC 2119, March 1997. [Online].
**[std:rfc3986] Tim Berners-Lee, Roy Fielding, and Larry Masinter.**
Uniform Resource Identifier (URI): Generic syntax. RFC 3986, January 2005. [Online].
**[std:skosref] Alistair Miles and Sean Bechhofer, editors.**
SKOS reference. W3C Recommendation, August 2009. [Online].
**[std:turtle] Dave Beckett.**
Turtle - terse RDF triple language. W3C Team Submission, January 2008. [Online].
**[std:ucd] Sébastien Derriere, Andrea Preite Martinez, and Roy Williams, editors.**
UCD (Unified Content Descriptor) — moving to UCD1+. IVOA Recommendation, 2004. [Online, cited February 2008].
**[std:voevent] Rob Seaman and Roy Williams, editors.**
Sky event reporting metadata (VOEvent). IVOA Recommendation, 2006. [Online].
**[std:w3cdtf] Misha Wolf and Charles Wicksteed.**
Date and time formats. W3C Note, September 1997. [Online].
**[std:z39.19] Guidelines for the construction, format and management of monolingual thesauri (ANSI/NISO Z39.19-2005).**
American National Standard, 2005. Closely corresponds to BS 8723:2005, parts 1 and 2, which replaces BS 5723; and to forthcoming ISO 25964, which replaces ISO 2788. [Online].